# Determination of the high-pressure crystal structure of BaWO$_4$ and PbWO$_4$

D. Errandonea,[1,*] J. Pellicer-Porres,[1] F. J. Manjón,[2] A. Segura,[1] Ch. Ferrer-Roca,[1]

R. S. Kumar,[3] O. Tschauner,[3] J. López-Solano,[4] P. Rodríguez-Hernández,[4] S. Radescu,[4]

A. Mujica,[4] A. Muñoz,[4] and G. Aquilanti[5]

[1] Dpto. Fís. Aplicada-ICMUV, Universitat de València, Edificio de Investigación,
C/ Dr. Moliner 50, 46100 Burjassot (Valencia), Spain

[2] Dpto. Fís. Aplicada, Universitat Politècnica de València, Cno. de Vera s/n, 46022 València, Spain

[3] High Pressure Science and Engineering Center, Department of Physics,
University of Nevada, 4505 Maryland Parkway, Las Vegas, Nevada 89154-4002, USA

[4] Dpto. Fís. Fundamental II, Universidad de La Laguna, La Laguna, Tenerife, Spain

[5] European Synchrotron Radiation Facility, BP 220, Grenoble, F-38043 France

**Abstract:** We report the results of both angle-dispersive x-ray diffraction and x-ray absorption near-edge structure studies in BaWO$_4$ and PbWO$_4$ at pressures of up to 56 GPa and 24 GPa, respectively. BaWO$_4$ is found to undergo a pressure-driven phase transition at 7.1 GPa from the tetragonal scheelite structure (which is stable under normal conditions) to the monoclinic fergusonite structure whereas the same transition takes place in PbWO$_4$ at 9 GPa. We observe a second transition to another monoclinic structure which we identify as that of the isostructural phases BaWO$_4$-II and PbWO$_4$-III (space group P2$_1$/$n$). We have also performed *ab initio* total-energy calculations which support the stability of this structure at high pressures in both compounds. The theoretical calculations further find that upon increase of pressure the scheelite phases become locally unstable and transform displacively into the fergusonite structure. The fergusonite structure is however metastable and can only occur if the transition to the P2$_1$/$n$ phases were kinetically inhibited. Our experiments in BaWO$_4$ indicate that it becomes amorphous beyond 47 GPa.

* Corresponding author; electronic mail: daniel.errandonea@uv.es, Tel.: (34) 96 354 3680, FAX: (34) 354 3146

**I. Introduction**

Research and development are underway for the implementation of new instrumentation in low-background particle-physics experiments. Event-type discrimination measurements could be possible in the near future by means of simultaneous measurements of a combination of phonon and scintillation signals from cryogenic phonon-scintillation detectors **[1]**. These detectors provide unique advantages in experiments searching for rare events; i.e. interactions with weakly-interactive massive particles **[2]**, double-beta decays **[3]**, and radioactive decays of very long-living isotopes **[4]**. Both $CaWO_4$ and $PbWO_4$ are promising materials for the next generation of cryogenic phonon-scintillation detectors **[2, 5, 6]**. This has motivated a renewed interest on the fundamental physical properties of the $AWO_4$ tungstates (with A = Ca, Sr, Ba, Pb, Eu) which under normal conditions crystallize in the tetragonal scheelite structure [space group (SG): $I4_1/a$, No. 88, and number of formula units per crystallographic cell Z = 4] **[7]**. The scheelite tungstates are in fact technologically important materials within a wider scope, having been used during the last years as solid-state scintillators **[8 - 10]** and in other optoelectronic devices **[11 - 13]**. A significant amount of research work on the structural behavior of $AWO_4$ compounds exists **[14 – 23]** and this corpus forms a solid background for understanding the main physical properties of these materials.

We have recently established by means of angle-dispersive x-ray diffraction (ADXRD) experiments, x-ray absorption near edge structure (XANES) measurements, and *ab initio* total-energy calculations **[20, 21]**, that upon compression $CaWO_4$ and $SrWO_4$ undergo a scheelite-to-fergusonite phase transition. In the present work we report a similar combined study in $BaWO_4$ and $PbWO_4$. This allows us to get a more complete picture of the structural behavior of the AWO$_4$ scheelite tungstates and improves our understanding of their physical properties.

The existence of high-pressure (HP) high-temperature (HT) phases in BaWO$_4$ and PbWO$_4$ has been known since the 1970s. The crystallographic structure of the HP-HT phase BaWO$_4$-II (SG: P2$_1$/*n*, No. 14, Z = 8) was resolved by Fujita *et al.* **[24]** and Kawada *et al.* **[25]**. The same structure was also found by Richter *et al.* **[26]** for the phase III of PbWO$_4$ previously discovered by Chang **[27]** (as in the case of BaWO$_4$-II, by combining high pressure and high temperature). For PbWO$_4$ another monoclinic phase, phase II (raspite, SG: P2$_1$/*a*, No 14, Z = 4), is known to occur in Nature as a metastable and minority form under normal conditions **[28]**. On the other hand, the occurrence of a pressure-driven phase transition at room temperature (RT) and 6.5 GPa in BaWO$_4$ and 4.5 GPa in PbWO$_4$ was observed in Raman experiments performed by Jayaraman *et al.* **[29, 30]**. These authors suggested that for both materials the high-pressure forms had the monoclinic HgWO$_4$-type structure (SG: C2/*c*, No. 15, Z = 4) **[31]**. However, in the single-crystal ADXRD experiments carried out by Hazen *et al.* in PbWO$_4$ no phase transition was observed up to 6 GPa **[19]**. In more recent high-pressure powder ADXRD experiments Panchal *et al.* observed that scheelite-BaWO$_4$ transforms to a high-pressure phase at 7 GPa and on the basis of the quality of the unit-cell fit these authors suggested that the structure of this phase could be fergusonite and not HgWO$_4$-type **[16]**. They also found evidence that beyond 14 GPa BaWO$_4$ undergoes another transition to an unidentified new phase. A single total-energy calculation has considered the possibility of the PbWO$_4$-III phase **[32]**. These facts show that despite the experimental and theoretical efforts made we have not yet achieved a full understanding of the effect of pressure on the structure of BaWO$_4$ and PbWO$_4$.

The goal of the present study is to examine comprehensively the crystal stability of BaWO$_4$ and PbWO$_4$ up to approximately 20 GPa. In order to improve the current understanding of the structural behavior of these compounds we have performed ADXRD and XANES measurements in a diamond-anvil cell (DAC) at RT as well as *ab initio* total-energy calculations on a number of phases. From our ADXRD measurements we find that both compounds undergo a scheelite-to-fergusonite phase transition (at 7.1 GPa in BaWO$_4$ and 9 GPa in PbWO$_4$). These transitions are supported by our high-pressure XANES measurements. In addition, we find that BaWO$_4$ and PbWO$_4$ undergo a second transition to the monoclinic BaWO$_4$-II and PbWO$_4$-III isostructural phases near 10 GPa and 15 GPa, respectively. The *ab initio* calculations find that the fergusonite phase can only occur as a metastable phase in both compounds and that the BaWO$_4$-II and PbWO$_4$-III phases are respectively stable at pressures above 7 GPa and 9 GPa. We think that the intermediate fergusonite phase was experimentally observed due to a kinetic hindrance of the $I4_1/a$-to-$P2_1/n$ transformation. We also find that amorphization occurs in BaWO$_4$ at pressures exceeding 47 GPa.

The details of the experimental methods are described in Section II and those of the *ab initio* theoretical calculations in Section III. Our experimental ADXRD results (including the evolution of the crystalline structures of both compounds under pressure) are given in Sec. IV.A, the results of the XANES study in Sec. IV.B, and the results of the theoretical study in Sec. IV.C. We summarize our conclusions in Sec. V.

**II. Experimental Details**

BaWO$_4$ and PbWO$_4$ crystals were grown with the Czochralski method starting from raw powders having 5N purity **[8, 11]**. Samples were prepared as fine ground powders from the single crystals. High-pressure ADXRD measurements were carried out in a 400 μm culet Mao-Bell DAC. Powder samples were loaded together with a ruby chip into a hole 100 μm in diameter drilled on a 40 μm thick rhenium gasket. For the XANES measurements under pressure, fine powder samples were loaded together with a ruby chip into a hole 200 μm in diameter drilled on a 50 μm thick Inconel gasket and inserted between the diamonds of a 400 μm culet membrane-type DAC. Silicone oil was used as pressure-transmitting medium in all the experiments. The pressure was measured by the shift of the R1 photoluminescence line of ruby [33].

The ADXRD data shown in the present paper are based on three independent runs on $BaWO_4$ (up to 9 GPa, 25 GPa and 56 GPa) and one run on $PbWO_4$ (up to 19.5 GPa). ADXRD experiments were performed at the 16-IDB beamline of the HPCAT facility at the Advanced Photon Source (APS). Monochromatic synchrotron radiation at $\lambda = 0.3679$ Å (on the $BaWO_4$ samples) or $\lambda = 0.3888$ Å (on the $PbWO_4$ samples) was used for data collection on a Mar345 image plate. The x-ray beam was focused down to 10 x 10 μm$^2$ using Kickpatrick-Baez mirrors. The diffraction images were integrated and corrected for distortions using FIT2D [34] to yield intensity versus 2θ diagrams. Indexing, structure solution, and refinements were performed using the GSAS [35] and POWDERCELL [36] program packages. XANES experiments were conducted at the ID24 energy-dispersive x-ray absorption station of the European Synchrotron Radiation Facility (ESRF) [37, 38]. Experiments were performed at the W $L_3$-edge (10.207 keV). A curved Si (111) monochromator [39] and a vertically focusing mirror defined a focus spot of 30 x 20 μm$^2$. The reference standard for the energy calibration was metallic W. A detailed description of the ADXRD and XANES experiments is given in Ref. [20].

**III. Details of the total-energy calculations**

Further to the ADXRD and XANES experiments the structural phase-stability of $BaWO_4$ and $PbWO_4$ was theoretically studied by means of total-energy calculations performed within the framework of the density functional theory (DFT) with the Vienna *ab initio* simulation package (VASP) **[40]**. The exchange and correlation energy was dealt within the generalized gradient approximation (GGA) **[41]**. A review of DFT-based total-energy methods as applied to the theoretical study of phase stability can be found in Ref. **[42]**. For the present calculations on PbWO$_4$ we used ultrasoft Vanderbilt-type pseudopotentials **[43]** while for BaWO$_4$ we adopted the projector augmented wave (PAW) scheme **[44]**. Both the semicore 5d electrons of Pb and the semicore 5s and 5p electrons of Ba were dealt with explicitly in the calculations. We used basis sets of plane waves up to a kinetic-energy cutoff of 692.5 eV for PbWO$_4$ and 875 eV for BaWO$_4$, and Monkhorst-Pack grids for the Brillouin-zone integrations which ensure highly converged and precise results (to about 1 meV per formula unit). At each selected volume the structure of the phases considered was relaxed through the calculation of the forces on the atoms and the components of the stress tensor, which in the equilibrium yielded the values of the internal and cell parameters of the structural phases. Various structural information (equilibrium volume, bulk moudulus, etc) for each phase was obtained from the calculated energy-volume curves after Birch-Murnaghan fitting.

**IV. Results and discussion**

**A. ADXRD measurements at high pressures**

**A.1. Low-pressure phase and phase transition**

The *in situ* ADXRD data measured at different pressures are shown in **Fig. 1(a)** for BaWO$_4$ and **Fig.1(b)** for PbWO$_4$. The x-ray patterns could be indexed within the scheelite structure (stable at normal conditions) up to 6.9 GPa in BaWO$_4$ and up to 8.1 GPa in PbWO$_4$. For BaWO4, splitting and broadening of the diffraction peaks are observed at 7.3 GPa together with the appearance of new reflections (depicted by arrows in **Fig. 1(a)**), in particular the weak peaks observed at $2\theta \approx 3.5º$ and at $2\theta \approx 9.5º$.

These facts are indicative of a structural phase transition around 7.1(2) GPa which is in agreement with previous observations [16, 29]. In the case of PbWO$_4$, a broadening of the Bragg peaks is observed together with the appearance of new reflections in the x-ray pattern measured at 10.1 GPa. The same low-angle reflections found in BaWO$_4$ and previously observed in the ADXRD patterns of the high-pressure phases of CaWO$_4$ and SrWO$_4$ [20] were also present in PbWO$_4$. In spite of the typical broadening of the diffraction peaks observed in all scheelite-type tungstates (independently of the pressure-transmitting medium used in the experiments) we are able to place the threshold of a structural phase transition in PbWO$_4$ at 9.1(10) GPa.

The value of 7.1(2) GPa for the transition pressure in BaWO$_4$ compares well with that of 6.5 GPa reported by Jayaraman *et al.* [29] in their Raman study. (Values of transition pressures obtained from Raman scattering tend to be slightly lower than those obtained by x-ray diffraction on the same material [45].) However, the observed pressure of 9.1(10) GPa for the phase transition in PbWO$_4$ is much larger than the value of 4.5 GPa reported in another previous Raman study by Jayaraman *et al.* on this material [30]. Using the same technique, the authors of Ref. [30] found a value of 9 GPa for the transition in PbMoO$_4$, a material that should have a similar behaviour than PbWO$_4$ on account of the similarities in their mean Pb-O distances and the WO$_4$/Pb and MoO$_4$/Pb radii ratios [18]. Application of the size criterion proposed in Ref. [18] to PbWO$_4$ leads to a transition pressure of 7.9(13) GPa which is close to our measured value. The fact that Hazen *et al.* [19] did not observe any phase-transition in their x-ray diffraction experiments performed up to 6 GPa further suggests that the transition pressure of PbWO$_4$ was probably underestimated in the Raman study of Ref. [30]. As we will see below, our XANES measurements and *ab initio* total-energy calculations further support this conclusion.

The pressure dependence of the lattice parameters, cell volume, and axial ratios for both BaWO$_4$ and PbWO$_4$ are shown in **Figs. 2(a)-(d)**, where we also compare with previous high-pressure measurements **[16, 19]** and data at ambient conditions **[7]**. The pressure-volume curves shown in **Fig. 2(c)** were obtained from fitting of the experimental data using a third-order Birch-Murnaghan equation of states (EOS). The bulk modulus, its pressure derivative, and the atomic volume at zero pressure obtained for the scheelite phase of BaWO$_4$ and PbWO$_4$ are summarized and compared with those of other scheelite tungstates in Table I. The values of these parameters are in good agreement with previously reported data **[16, 19]** and indicate that BaWO$_4$ is the most compressible scheelite-like orthotungstate. This is expected since the average Ba-O distance at ambient conditions (2.768 Å) is larger than the average A-O distance ($d_{A-O}$) in any of the other AWO$_4$ compounds, and the bulk modulus is known to be proportional to $(d_{A-O})^{-3}$ **[20]**.

**Figures 2(a), 2(b)**, and **2(d)** show that for the scheelite phases the compressibility along the *c*-axis is larger than that along the *a*-axis. This observation extends to CaWO$_4$, SrWO$_4$, and EuWO$_4$ **[20, 22]** and can be related to the fact that when pressure is applied to the scheelite structure the WO$_4$ tetrahedra remain essentially undistorted while the volume of the AO$_8$ bisdisphenoids is largely reduced. Thus, the *a*-axis is expected to be less compressible than the *c*-axis because the WO$_4$ units are directly aligned along the *a*-axis whereas along the *c*-axis there is an A-cation between the WO$_4$ tetrahedra (see **Fig. 1** of Ref. **[20]**). From the present ADXRD results we further obtain that for both scheelite-BaWO$_4$ and scheelite-PbWO$_4$ the W-O distances are more rigid than the Ba-O and Pb-O distances (see **Fig. 3**). In particular, the W-O distance in BaWO$_4$ decreases from 1.838 Å at 1 bar to 1.785 Å at 6.9 GPa and the average Ba-O distance decreases from 2.768 Å at 1 bar to 2.653 Å at 6.9 GPa. Similar

differences between the changes of the W-O and Pb-O distances upon compression are observed in PbWO$_4$. These features are consistent with the previous argument about the anisotropic compressibility of the AWO$_4$ compounds [20].

### A.2. High-pressure phases

The ADXRD spectra of BaWO$_4$ exhibit a change around 7.3 GPa, while in the spectra of PbWO$_4$ a similar change occurs near 10 GPa, see **Figs. 1(a)** and **1(b)**. These changes are completely reversible upon pressure release but significant hysteresis was observed. In both compounds the increase of pressure above the threshold of the transition leads to an increase of the broadening of the Bragg peaks. Splitting of the Bragg peaks is clearly observed in BaWO$_4$ above the phase transition [cf. the ADXRD patterns in **Fig. 1(a)** collected at 7.3 GPa and 10.5 GPa: the (200) and (002) peaks located near 2θ ≈ 8º split, as well as the (240) and (042) peaks near 2θ ≈ 10.5º and the ($\bar{2}$02) and (202) peaks near 2θ ≈ 11º]. The same splitting is not clearly observed in PbWO$_4$ because of the broadening of the diffraction peaks. We note that the broadening and splitting of Bragg peaks in BaWO$_4$ is observed because of the increase of the monoclinic distortion: both the monoclinic angle β and the difference between the b/a and b/c axial ratios increases. In particular, β goes from 90.08º at 7.3 GPa to 90.98º at 10.5 GPa.

**Figures 1(b)** and **4** show the LeBail refinement [46] of the ADXRD pattern of PbWO$_4$ measured at 10.1 GPa and the Rietveld refinement of the experimental spectra of BaWO$_4$ measured at 7.3 GPa assuming the fergusonite structure. A good fitting is obtained for both tungstates, with residuals $R_{WP}$ = 1.9%, $R_P$ = 1.6%, and $R(F^2)$ = 1.2 % for BaWO$_4$ at 7.3 GPa (242 reflections) and $R_{wp}$ = 2.35%, $R_P$ = 2.2%, $R(F^2)$ = 1.8% for PbWO$_4$ at 10.1 GPa (210 reflections). The new characteristic peaks of the fergusonite structure observed for CaWO$_4$ and SrWO$_4$ [20] are here also observed in PbWO$_4$ close to $2\theta = 4°$ and $2\theta = 10°$ at 10.1 GPa and in BaWO4 close to $2\theta = 3.5°$ and $2\theta = 9.5°$ at 7.3 GPa. Table II summarizes the lattice parameters and atomic positions of BaWO$_4$ at 1 GPa and 7.3 GPa, and of PbWO$_4$ at 0.7 GPa. The lattice parameters of fergusonite-PbWO$_4$ at 10.1 GPa are: $a = 5.376(6)$ Å, $b = 11.495(5)$ Å, $c = 5.273(7)$ Å, and $\beta = 91.57(9)°$. The atomic positions of fergusonite-PbWO$_4$ at 10.1 GPa cannot be extracted from the experimental data because of the impossibility of performing a Rietveld refinement of the spectrum. The parameters obtained for fergusonite-BaWO$_4$ at 7.3 GPa [$a = 5.465(7)$ Å, $b = 12.109(3)$ Å, $c = 5.439(7)$ Å, $\beta = 90.087(9)°$] agree reasonably well with those reported by Panchal *et al.* at 8.3 GPa ($a = 5.444$ Å, $b = 12.290$ Å, $c = 5.236$ Å, $\beta = 89.56°$) **[16]**. Our high-resolution synchrotron ADXRD experiments allowed us a large access to the reciprocal space and the possibility of performing Rietveld refinements, while Panchal *et al.* **[16]** performed quite valuable ADXRD measurements using a rotating anode machine that only allowed them to perform Le Bail refinements with a lower resolution and smaller access to the reciprocal space than ours. Panchal *et al.* **[16]** also concluded that the high-pressure phase of BaWO$_4$ is fergusonite but with a larger uncertainty in the lattice parameters.

In BaWO$_4$ evidence of the occurrence of a second phase transition is observed at 10.9 GPa. At this pressure we observed an extra broadening of the diffraction peaks, accompanied by several changes in the diffraction pattern [see **Fig. 1(a)**]. Upon further increase of pressure the broadening of the peaks further increases, however at least up to 15.2 GPa all the ADXRD patterns can be assigned to the same phase observed at 10.9 GPa. Evidence of the occurrence of a second pressure-induced phase transition was previously observed by Panchal *et al.* **[16]** albeit at the larger pressure of 14 GPa. However, these authors did not report any x-ray diffraction pattern between 9.3 and 14 GPa. The results of our XANES study and *ab initio* calculations (see below) further locate the occurrence of a second transition around 10 GPa in BaWO$_4$. We think that the onset of this transition was previously overestimated in Ref. **[16]**.

The quality of the ADXRD patterns collected from the new phase of BaWO$_4$ did not allow a Rietveld refinement. Therefore, we analyzed them using the Le Bail extraction technique considering the following candidate structures: raspite, HgWO$_4$, BaWO$_4$-II, wolframite (SG: P2/*c*, No. 13, Z = 2) **[47]**, α-MnMoO$_4$ (SG: C2/*m*, No. 12, Z = 8) **[15]**, LaTaO$_4$ (SG: P2$_1$/*c*, No. 14, Z = 4) **[48]**, BaMnF$_4$ (SG: A2$_1$/*am*, No. 36, Z = 4) **[49]**, SrUO$_4$ (SG: P*bcm*, No. 57, Z = 4) **[50]**, C*mca* (SG: C*mca*, No. 64, Z = 8) **[20]**, zircon (SG: I4$_1$/*amd*, No. 141, Z = 4) **[51]**, and pseudo-scheelite (SG: P*nma*, No. 62, Z = 4) **[52]**. The most likely space group was found to be P2$_1$/*n* corresponding to the BaWO$_4$-II-type structure. In the spectrum collected at 10.9 GPa the Le Bail extraction technique converged to the structure of BaWO$_4$-II with residuals R$_{WP}$ = 2.2%, R$_P$ = 1.6%, and R(F$^2$) =1.2% for 909 reflections. The lattice parameters obtained for this structure at 10.9 GPa are $a$ = 12.841(9) Å, $b$ = 7.076(6) Å, $c$ = 7.407(6) Å, $\beta$ = 93.0(9)°. **Fig. 4** shows the refined structure models and the residual of the refinement procedure. All the ADXRD patterns measured up to 15.2 GPa can also be assigned to the BaWO$_4$-II-type structure. From our data we conclude that the transition from the fergusonite phase to the BaWO$_4$-II phase occurs at 10.7(2) GPa together with a large volume collapse ΔV/V=8%. This volume collapse reflects the fact that the structure of BaWO$_4$-II consists of densely packed networks of distorted WO$_6$ octahedra. We note here that although we propose the identification of the new phase as BaWO$_4$-II, the LaTaO$_4$-type and the BaMnF$_4$-type structures were in close competence with it giving only slightly larger residuals in our analysis. Peak intensities can be slightly distorted in DAC experiments **[53, 54]** and thus an analysis based on residuals alone is not enough in this case to discriminate between the crystal structures. However, both the BaMnF$_4$-type and LaTaO$_4$-type structures fail to account for the presence of the peaks located at 2θ ≈ 3.3º, 4.1º, and 5.5º [assigned to the (200), (011), and (211) Bragg peaks of the BaWO$_4$-II structure, respectively].

The x-ray patterns of BaWO$_4$ measured beyond 15.2 GPa keep slowly changing up to 47 GPa. In particular, the continuous broadening of the Bragg peaks observed from 15.2 GPa to 47 GPa suggests that beyond 15.2 GPa there is a substantial increase of the disorder in the crystalline phase. At 56 GPa all the peaks finally disappear under a broad diffuse scattering. We think that these two facts indicate that at such high compression BaWO$_4$ amorphizes, similarly to what has been observed in CaWO$_4$ **[14]**. The occurrence of this kind of pressure-induced amorphization is sometimes related to a frustrated solid-solid phase transition. In the present study all the changes observed up to 25 GPa were reversible although showing significant hysteresis. However, the changes observed beyond 47 GPa were *not* reversible, a fact which is consistent with pressure-induced amorphization. Another possibility for the appearance of the broad features in the ADXRD patterns is the occurrence of a pressure-induced chemical decomposition of BaWO$_4$ as observed in LiGdF$_4$ **[55]**. This is however not likely the case for CaWO$_4$ since annealing at 45 GPa and 477 K during two hours led to the nucleation of a new crystalline structure of CaWO$_4$ **[15]**. Most likely the lost of the Bragg peaks in the ADXRD patterns of BaWO$_4$ and CaWO$_4$ (near 47 GPa and 40 GPa, respectively) is the result of the frustrated transformation of their high-pressure crystalline phases into a non-crystalline solid. The irreversible nature of the amorphization implies that beyond 47 GPa the polyhedra not only deform and interconnect probably differently, but also that the structural changes are significantly larger to hinder the reversal of deformations upon release of pressure. Amorphization can be understood in terms of the packing of the anionic WO$_4$ units around the A cations (size criterion) **[18]**. When the ionic radii of the $WO_4$ groups is small relative to that of the A cations, increasing repulsive and steric stresses induced by pressure can be accommodated by deformation of the cation outer shell as opposed to significant changes in its average position, thereby favouring the transformation to a high-pressure crystalline phase. In contrast to this, if the ratio between the ionic radii ($WO_4$/A) is large the material will accommodate increased stresses through larger and more varied displacements from their average positions resulting in a subsequent loss of translational periodicity at high pressure. The lower pressure for the onset of amorphization in $CaWO_4$ as compared to $BaWO_4$ is consistent with the larger $WO_4$/A ratio for the Ca compound ($WO_4$/Ca = 1.89 and $WO_4$/Ba = 1.47). A direct conclusion can be drawn for $SrWO_4$ and $PbWO_4$ for which the ratios $WO_4$/Sr = 1.76 and $WO_4$/Pb = 1.66 imply that they should amorphize at pressures around 43 GPa and 45 GPa, respectively.

In the case of $PbWO_4$ the broadening of the diffraction peaks continuously increases from the transition pressure of 9.1 GPa up to 19.5 GPa, the maximum pressure achieved in our experiments on $PbWO_4$. One of the possible reasons for the broadening of the Bragg peaks of the fergusonite structure is the continuous increase of the monoclinic distortion. At 13.6 GPa, the splitting and broadening of the Bragg peaks has increased considerably, but the ADXRD pattern can still be assigned to a strongly distorted fergusonite structure. In particular the low-angle peaks are still present. However, we have also observed some reversible changes on the features of the diffraction pattern at 15.6 GPa. The intensity of the pattern decreases, the two weak peaks at low 2θ angles apparently disappears, and the strong peaks observed near 11.5º at 10.1 GPa shifts to lower 2θ angles. The patterns collected above 15.6 GPa show the same features than the one collected at 15.6 GPa. The observed changes may relate to the occurrence of a structural change in $PbWO_4$ at 15.6 GPa. As we will show below, this hypothesis is confirmed by the present XANES results. We think that the continuous broadening observed in the ADXRD patterns of PbWO$_4$ beyond 10.1 GPa may be caused by an increase of the disorder in the crystalline structure of PbWO$_4$. This result suggests that perhaps PbWO$_4$ will also amorphize at a pressure higher than the maximum pressure reached in our experiments, as it occurs in BaWO$_4$ and CaWO$_4$. The fact that the broadening observed up to 19.5 GPa is reversible (see **Fig. 1(b)**) would then indicate that the pressure-induced structural disorder did not develop enough at 19.5 GPa to avoid the reversion to the original structure upon decompression.

**B. XANES measurements at high pressures**

XANES measurements provide information about the geometrical arrangement of the atoms surrounding the absorbing atom. XANES spectra do not depend on long range order, so the information that they provide is complementary to the one yielded by x-ray diffraction. As a sensitive tool XANES can be used to obtain information about pressure-driven structural changes. For the present study we have carried out high-pressure XANES experiments on BaWO$_4$ and PbWO$_4$ in order to study the evolution of the W environment under compression. We use XANES spectra as a footprint to help identifying the occurrence of high-pressures phases. Specifically, we compare experimental spectra with those calculated for a given phase using the real-space multiple-scattering code implemented in the FEFF8 package **[56]**.

The XANES simulations use as input the structural information provided by *ab initio* calculations or x-ray diffraction. The self-consistent potential was calculated using 6.5 Å clusters and the Hedin-Lundqvist energy-dependent self-energy. Full multiple scattering calculations were performed with the same cluster size. No pseudo Debye-Waller factor has been considered in our simulations.

We have performed simulations considering several structures, including fergusonite, wolframite, LaTaO$_4$, BaMnF$_4$, HgWO$_4$, C*mca* and BaWO$_4$-II (or PbWO$_4$-III), see **Fig. 5**. The structural data used were taken from the literature in the case of the HgWO$_4$ **[31]**, wolframite **[47]**, and PbWO$_4$-III structures **[26]**. For the rest of the structures considered they were obtained either from the present Rietveld refinements (see Table II) or from our *ab initio* calculations (see Table III)**.** As previously observed in CaWO$_4$ and SrWO$_4$ **[20]**, the B resonance is very sensitive to the W-O coordination. In those structures where W is fourfold coordinated by O atoms, as the scheelite and the fergusonite structures, the B resonance is clearly visible. On the other hand, when W atoms approach a sixfold coordination (wolframite, LaTaO$_4$, BaMnF$_4$, HgWO4, and BaWO$_4$-II) the B resonance disappears. Differences between the spectra corresponding to structures where W is sixfold coordinated by O atoms are more subtle.

**B.1. Low-pressure phase**

Experimental XANES spectra of BaWO$_4$ and PbWO$_4$ at the W L$_3$-edge (10.207 keV) are shown in **Fig 6**. The scheelite spectra of both compounds are quantitatively similar, with five evident resonances. Similar features were present in a previous study on CaWO$_4$ and SrWO$_4$ **[20]**. We find a tendency in the AWO$_4$ series that the intensity of the E resonance slightly increases whereas the resonances B, C, and D become less pronounced as the A cation becomes heavier. The calculated position and relative intensity of the resonances (**Fig. 5**) agrees qualitatively with the experimental spectra. (The resonances in the theoretical spectra are more pronounced because the pseudo Debye-Waller factor has not been considered in the simulations.)

**B.2. High-pressure phases**

The high-pressure XANES spectra of BaWO$_4$ remain essentially unchanged up to 7.8 GPa. At this pressure the B and C resonances loose intensity, whereas the white line (A) starts to broaden. These changes stabilize at 9.8 GPa, at which pressure the B resonance has disappeared. In the downstroke the scheelite phase is recovered, although with significant hysteresis.

We interpret the changes observed at 7.8 GPa as the beginning of a change in the W-O coordination from four to six, which is completed at 9.8 GPa. With respect to the possible candidates to the high-pressure phase observed beyond 9.8 GPa, the observation of resonances F and G, as well as their relative intensity and energy points to the $BaWO_4$-II, $BaMnF_4$-type, and $LaTaO_4$-type structures. On the other hand, in the XANES simulations corresponding to the $BaMnF_4$ and $LaTaO_4$ structures a weak resonance H is present, but it is not clearly evidenced in the experimental spectra. Thus, we conclude that from the XANES point of view the $BaWO_4$-II structure is the most convincing candidates for the crystalline structure of $BaWO_4$ beyond 9.8 GPa.

Apparent discrepancies arise here between XANES and ADXRD (and Raman) experiments. According to the present and previous ADXRD experiments **[15]** the structural sequence of $BaWO_4$ is scheelite $\rightarrow$ ferguosonite $\rightarrow$ $BaWO_4$-II. The first transition occurs at around 7 GPa according to Raman and ADXRD measurements, and the second one occurs around 11 GPa. However, our XANES measurements cannot confirm the existence of the ferguosonite phase and suggests the presence of the $BaWO_4$-II structure at 9.8 GPa. There is a fact that can be argued against these apparent discrepancies. The distortion of the W-O tetrahedra at the scheelite-to-ferguosonite transition is subtle. In fact, in $AWO_4$ scheelite compounds the transition is caused by displacements of the A cations from their high-symmetry positions. Therefore, the XANES spectrum of the ferguosonite phase should not differ much from that of the scheelite phase (see **Fig. 5** and Ref. **[20]**). In fact only small changes of intensities in the resonances B, C, D, and E are expected. As we mentioned above, these resonances become less pronounced as the A cation atomic number increases. This fact makes the transition more difficult to detect in $BaWO_4$ than in $CaWO_4$ and $SrWO_4$. In fact, in Ref. **[20]** the XANES measurements detected the transition in $CaWO_4$ at the *same* pressure than the ADXRD measurements, but in $SrWO_4$ XANES detected the transition at *higher* pressures than ADXRD; only after the monoclinic distortion is enhanced under compression. In $BaWO_4$ a second transition occurs quite close to the scheelite-to-fergusonite transition (before the monoclinic distortion is enhanced). We think that probably because of all these reasons the presence of the fergusonite phase was not detected in our XANES experiments.

As regards to $PbWO_4$, the spectra in **Fig. 6b** show a decrease in the intensity of the B resonance starting at 9.0 GPa. At 10.9 GPa the B resonance fades out and new weak resonances J, K and L appear. The absence of B resonance in the high-pressure phase seems to contradict ADXRD measurements, since ADXRD indicates a scheelite-fergusonite transition near 9 GPa and, according to the trend followed by simulations and experiments in $CaWO_4$ and $SrWO_4$, as well as the result of fergusonite simulations in $PbWO_4$, resonance B should not disappear. The apparent discrepancy is solved thanks to the *ab initio* computations. Its results will be fully discussed in the next section. Now we would like only to point out that, following the calculations, the internal parameters of the fergusonite structure change between 7.9 GPa and 9.5 GPa, implying a change in the W coordination from 4 oxygen atoms to 4+2 (see **Fig. 3**). This means that in $PbWO_4$ the fergusonite structure would play the role of bridge phase between the scheelite structure, composed of $WO_4$ tetrahedra, and a high-pressure structure containing $WO_6$ octahedra. The FEFF simulations carried out with the two sets of internal parameters, and thus the two different coordinations, are shown in **Fig. 5b**. As expected, in the fourfold coordinated version of the fergusonite structure the B resonance is present, whereas in the sixfold coordinated version the B resonance is absent. We thus conclude that at 9 GPa a phase transition takes place in PbWO$_4$ towards a ferguosonite structure, being upon further compression the W environment distorted in such a way that the W atoms are surrounded by six O atoms in a 4+2 configuration at 10.9 GPa.

At 16.7 GPa there are new, subtle changes in the spectra. Resonances J and L loose intensity, whereas resonance K becomes enhanced. These changes are compatible with a second phase-transition towards the PbWO$_4$-III or LaTaO$_4$ structures. The occurrence of this second transition is not only in agreement with our ADXRD experiments, but also with recent Raman [57] and optical absorption measurements [58] which detected the occurrence of two phase transitions located at 7 GPa and 12 GPa. The absence of the weak M resonance in the experimental spectra is not a definitive argument but suggests that the PbWO$_4$-III phase is the more likely candidate for the third high-pressure phase.

To close this section we would like to comment that in PbWO$_4$ the spectra collected at 3.4 GPa in a sample recovered from 24.1 GPa differs from that of the scheelite phase. However, when releasing the pressure from 19.5 GPa the scheelite structure was recovered at ambient pressure in our ADXRD experiments. The origin of this discrepancy remains unknown.

**C. Ab initio calculations at high pressures**

We turn now our attention to the results of our *ab initio* study on the energetics of the structural phases of BaWO$_4$ and PbWO$_4$ and the comparison with the experimental data reported in the previous sections. On account of either its present observation or its previous consideration within the family of scheelite compounds to which BaWO$_4$ and PbWO$_4$ belong, we have considered the following structures in our calculations: the scheelite structure itself, raspite (observed in PbWO$_4$ as a metastable phase under normal conditions), fergusonite, the BaWO$_4$-II-type (or PbWO$_4$-III-type) quenched from HP-HT conditions in both compounds, wolframite, HgWO$_4$-type, LaTaO$_4$-type, BaMnF$_4$-type, SrUO$_4$-type, zircon, and the orthorhombic C*mca* structure previously proposed by us in Ref. [20] as a very high-pressure phase in SrWO$_4$ and CaWO$_4$. We will comment only on the most relevant aspects of the theoretical results.

**Figure 7** shows the energy-volume curves obtained for each of the structures considered in BaWO$_4$ and PbWO$_4$. The relative stability and coexistence pressures of the phases can be obtained from these curves by the common-tangent construction [42]. **Figure 7** shows the scheelite phases of both compounds as being stable at zero and low pressure, with values of the volume per formula unit (pfu), bulk modulus and pressure-derivative of the bulk modulus in the equilibrium of, respectively, $V_0$= 105.2 Å$^3$, $B_0$= 52 GPa, and $B_0$'= 5 for BaWO$_4$ and $V_0$= 94.0 Å$^3$, $B_0$= 66 GPa, and $B_0$'= 4.7 for PbWO$_4$. These values are in good agreement with our experimental results, see Table I. Our calculated values of the structural parameters for scheelite-BaWO$_4$ [O(16f) at (0.2315, 0.1234, 0.0461) and *c/a*=2.25 at 0.9 GPa] and for scheelite-PbWO$_4$ [O(16f) at (0.2334, 0.1114, 0.0436) and *c/a*=2.21 at 0 GPa] are also in good agreement with the experimental values, see Table II.

For PbWO$_4$ we find the raspite structure to be very close in energy and equilibrium volume to the scheelite structure, with raspite slightly higher by about 20 meV pfu, which is in perfect agreement with raspite-PbWO$_4$ (or PbWO$_4$-II) being found in Nature as a metastable form under normal conditions [28]. This result disagrees with a previous theoretical calculation that obtained the raspite form lower in energy than the scheelite structure [32] (A similar disagreement exists for the PbWO$_4$-III phase which in Ref.[32] is shown much lower in energy than the scheelite form.). However in our study the raspite structure is placed significantly higher in energy in $BaWO_4$, for which such structure has not been reported experimentally: from our present results this form is not expected to occur in this material.

As pressure increases scheelite-$BaWO_4$ becomes unstable against $BaWO_4$-II and similarly scheelite-$PbWO_4$ becomes unstable to $PbWO_4$-III (isomorphous to $BaWO_4$-II). The unit-cell parameters and atomic positions of both phases are given in Table III. The calculated values of the coexistence pressures between both structures are about 5 GPa for both materials which is smaller than the experimental transition pressure at room temperature. It is however very likely that at RT kinetic barriers preclude the transition at such pressure: the first observations of these phases required heating to 400-600ºC as well as application of high pressure **[24-27]**.

At pressures around 7.5 GPa in $BaWO_4$ and 8 GPa in $PbWO_4$ the scheelite phases of these compounds become locally unstable and undergo a distortion into the fergusonite structure with none or very little volume collapse. At such pressures however the stable phase (the one with lower enthalpy) is the $P2_1/n$ structure of $BaWO_4$-II and $PbWO_4$-III. From our calculations there is thus no true range of stability for the fergusonite phase in either compound, and this can exist only as a metastable phase in the case that the transition to the monoclinic structure with space group $P2_1/n$ were inhibited. The fergusonite distortion of the scheelite structure is particularly small both structurally and energetically in $BaWO_4$ and thus the energy-volume curves for these phases are not distinguishable in **Fig.7.**

The transitions from the low-pressure scheelite phases to the high-pressure $P2_1/n$ phases ($BaWO_4$-II and $PbWO_4$-III) are first-order reconstructive transformations with a large associated change in volume (about 12% for $BaWO_4$ and 9% for $PbWO_4$, in good agreement with the experiments **[25, 26]**) and involve extensive rearrangement of the crystal structure. Because of this, the rate at which these transformations occur may be very slow and unstable (metastable) polymorphs may exist for very long periods of time at room temperature. This may be in the origin of the observation of the ferguson structure in the ADXRD experiments under compression. The fact that $PbWO_4$-III and $BaWO_4$-II exists at high pressure above 350 ºC and 600 °C, and are quenchable to ambient conditions **[25, 26]** supports the idea that in the experiments ferguson is observed instead of $PbWO_4$-III ($BaWO_4$-II) due to a kinetic hindrance of the equilibrium phase transformation. On the other hand, the scheelite-to-ferguson transition is a displacive transformation involving only small adjustments to the crystal structure. Such barrier-less continuous or cuasi-continouous transformations with none or very little change in volume are fully reversible, as observed in the experiments.

On further pressure increase in $BaWO_4$-II we find that this phase becomes unstable to the $BaMnF_4$-type structure around 27 GPa. (This $BaMnF_4$-type structure is related to the $LaTaO_4$-type structure, which turns out to be rather close in energy, and we find that one transforms continuously into the other as the compression progresses – such transition has indeed been reported as induced by temperature in the case of the compound $LaTaO_4$ itself **[59]**.) On further increase of pressure we find that the orthorhombic C*mca* structure (or *silvanite*) that we proposed in a previous study for $CaWO_4$ and $SrWO_4$ **[20]** becomes favored above around 56 GPa. In the case of $PbWO_4$ the C*mca* structure becomes favored over the $P2_1/n$ phase ($PbWO_4$-III) around 35 GPa.

Now, we would like to discuss the evolution of the ferguson phase of $PbWO_4$ under compression. From our calculations we obtain that, due to the change of the internal parameters, in this structure the W-O coordination changes gradually from 4 to 4+2. **Figure 3** gives the evolution obtained for the Pb-O and W-O bond distances of $PbWO_4$ under compression. There it can be clearly seen that theoretical calculations predict a change in the W coordination for PbWO$_4$ as already comment when discussing the XANES results. No similar evolution in the internal parameters exists in CaWO$_4$, SrWO$_4$ or BaWO$_4$; however, a similar evolution of the internal parameters has been found in YLiF$_4$ according to recent *ab initio* calculations **[60]**.

An interesting further result of our study is the fact that at expanded volumes (and corresponding *negative* pressures) the zircon structure becomes stable against the scheelite structure in both materials. Mineral zircon (ZrSiO$_4$) transforms under pressure to a structural phase isomorphous with scheelite **[61]** which is in agreement with our findings for BaWO$_4$ and PbWO$_4$ (though here of course the coexistence pressure is negative). Thus, including the expanded region (negative pressures) in the picture the following systematics arise for the first steps of the structural sequence undergone by these materials upon pressure increase: I4$_1$/*amd* (zircon) → I4$_1$/*a* (scheelite) → I2/*a* (fergusonite) → P2$_1$/*n*. The zircon structure transforms by means of a *translationgleiche* transition into the scheelite structure [twining zircon on (200), (020), and (002) generates the scheelite-type]. This structure is transformed into the fergusonite structure by means of another *translationgleiche* transition that involves a lowering of the point-group symmetry from 4/m to 2/m. And finally the P2$_1$/*n* structure is naturally obtained by a *klassengleiche* transition from fergusonite. This systematics may have important implications for a number of ABX$_4$ structures, specially those with a large difference between the sizes of the A and B atoms, which include some important minerals in addition to zircon and scheelite.

In order to better understand the structural behavior of ABX$_4$ structures under compression a parallel can be drawn between high-pressure transformations in the ABO$_4$ and MO$_2$ (MMO$_4$) octahedral structures. The rutile (MO$_2$) structure consists of infinite rectilinear rods of edge-sharing MO$_6$ octahedra parallel to *c*-axis, united by corner sharing to the octahedra in identical corner rods. If M is alternate substituted by bigger and smaller cations A and B and the cations are shifted in each rod then the zircon structure is obtained. Under compression rutile transforms to the α-PbO$_2$-type structure **[62]** and both scheelite and fergusonite can be thought as distorted superstructures of α-PbO$_2$. Additionally, the monoclinic post-fergusonite structure that we observed in BaWO$_4$ and PbWO$_4$ is related to baddeleyite, post-α-PbO$_2$ structure of rutile-type TiO$_2$ **[63]**. Thus the crystal chemistry systematics of MO$_2$ compounds provides additional support to the structural sequence that we extract for ABO$_4$ compounds from our results.

**V. Concluding Remarks**

We have studied the high-pressure behavior of BaWO$_4$ and PbWO$_4$ and have found that both compounds undergo a scheelite-to-fergusonite phase transition. On further increase of pressure both compounds transform to a monoclinic structure with space group symmetry P2$_1$/*n*. According to our *ab initio* calculations the scheelite structure is predicted to transform under compression directly to P2$_1$/*n*. We attribute the occurrence of the fergusonite phase in the sequence of structural transitions to the existence of a kinetic barrier that prevents the I4$_1$/*a*-to-P2$_1$/*n* for taking place. These conclusions may have important geophysical and geochemical implication since the scheelite-structured orthotungstates are common accessory minerals in various kinds of rocks in the Earth´s upper mantle. Pressures of 7 to 10 GPa and temperatures higher than 500 ºC are found at a depth of < 100 Km in the upper mantle **[64]**. Therefore, this family of minerals that in the Earth´s surface are isomorphous to scheelite are likely expected to be isostructural to BaWO$_4$-II in the Earth's upper mantle. Finally, we have found that further increase of pressure leads to amorphization of BaWO$_4$ in a similar way to that found previously for CaWO$_4$ **[14]**.

**Acknowledgments**

The authors thank Dr. P. Bohacek (Institute of Physics, Prague) and Dr. P. Lecoq (CERN) for providing the BaWO$_4$ and PbWO$_4$ crystals used to perform the experiments. This study was made possible through financial support from the Spanish government MCYT under grants numbers MAT2002-04539-CO2-01 and -02, and MAT2004-05867-C03-03 an -01. The U.S. Department of Energy, Office of Science, and Office of Basic Energy Sciences supported the use of the APS under Contract No. W-31-109-Eng-38. DOE-BES, DOE-NNSA, NSF, DOD-TACOM, and the Keck Foundation supported the use of the HPCAT. We would like to thank D. Häusermann and the staff at the HPCAT for their contribution to the success of the ADXRD experiments. XANES experiments have been done under proposal number HS-2412 at the ESRF. D. Errandonea acknowledges the financial support from the MCYT of Spain and the Universitat of València through the "Ramón y Cajal" program. A. Muñoz acknowledges the financial support from the Gobierno Autónomo de Canarias PI2003/174. J. López-Solano acknowledges the financial support from the Consejería de Educación del Gobierno Autónomo Canario. We also gratefully acknowledge the access to the computational resources of Mare Nostrum at the Barcelona Supercomputer Center.

**References**

[1] M. Bravin, M. Bruckmayer, C. Bucci, S. Cooper, S. Giordano, F. von Feilitzsch, J. Hohne, J. Jochum, V. Jorgens, R. Keeling, H. Kraus, M. Loidl, J. Lush, J. Macallister, J. Marchese, O. Meier, P. Meunier, U. Nagel, T. Nussle, F. Probst, Y. Ramachers, H. Sarsa, J. Schnagl, W. Seidel, I. Sergeyev, M. Sisti, L. Stodolsky, S. Uchaikin, and L. Zerle, Astroparticle Physics **12**, 107 (1999).

[2] G. Angloher, C. Bucci, C. Cozzini, F. von Feilitzsch, T. Frank, D. Hauff, S. Henry, T. Jagemann, J. Jochum, H. Kraus, B. Majorovits, J. Ninkovic, F. Petricca, F. Probst, Y. Ramachers, W. Rau, W. Seidel, M. Stark, S. Uchaikin, L. Stodolsky, and H. Wulandari, Nucl. Instrum. Methods Phys. Res. A **520**, 108 (2004).

[3] A. Alessandrello, V. Bashkirov, C. Brofferio, C. Bucci, D. V. Camin, O. Cremonesi, E. Fiorini, G. Gervasio, A. Giuliani, A. Nucciotti, M. Pavan, G. Pessina, E. Previtali, L. Zanotti, Phys. Lett. B **420,** 109 (1998).

[4] C. Cozzini, G. Angloher, C. Bucci, F. von Feilitzsch, D. Hauff, S. Henry, T. Jagemann, J. Jochum, H. Kraus, B. Majorovits, V. Mikhailik, J. Ninkovic, F. Petricca, W. Potzel, F. Probst, Y. Ramachers, W. Rau, M. Razeti, W. Seidel, M. Stark, L. Stodolsky, A. J. B. Tolhurst, W. Westphal, H. Wulandari, Phys. Rev. C **70**, 064606 (2004).

[5] G. Angloher, M. Bruckmayer, C. Bucci, M. Bühler, S. Cooper, C. Cozzini, P. Di Stefano, F. V. Feilitzsch, T. Frank, D. Hauff, T. Jagemann, J. Jochum, V. Jorgens, R. Keeling, H. Kraus, M. Loidl, J. Marchese, O. Meier, U. Ángel, F. Pröbst, Y. Ramachers, A. Rulofs, J. Schnagl, W. Seidel, I. Sergeyev, M. Sisti, M. Stark, S. Uchaikin, L. Stodolsky, H. Wulandari, and L. Zerle, Astroparticle Physics **18**, 43 (2002).

[6] S. Cebrian, N. Coron, G. Dambier, E. García, I. G. Irastorza, J. Leblanc, P. de Marcillac, A. Morales, J. Morales, A. Ortiz de Solórzano, J. Puimedon, M. L. Sarsa, and J. A. Villar, Astroparticle Physics **21**, 23 (2004).

[7] A. W. Sleigth, Acta Cryst. B **28**, 2899 (1972).

[8] A. A. Annenkov, M. V. Korzhik, and P. Lecoq, Nucl. Instrum. Methods Phys. Res. A **490**, 30 (2002).

[9] Compact Muon Solenoid (CMS), Technical Proposal, CERN/LHC 93-98, p.1 (1994).

[10] M. Kobayashi, M. Ishi, Y. Usuki, H. Yahagi, Nucl. Instrum. Methods Phys. Res. A **333**, 429 (1993).

[11] M. Nikl, P. Bohacek, N. Mihokova, M. Kobayashi, M. Ishii, Y. Usuki, V. Babin, A. Stolovich, S. Zazubovich, and M. Bacci, J. of Luminescence **87-89**, 1136 (2000).

[12] M. Nikl, P. Bohacek, N. Mihokova, N. Solovieva, A. Vedda, M. Martini, G. P. Pazzi, P. Fabeni, M. Kobayashi, M. Ishii, J. Appl. Phys. **91**, 5041 (2002).

[13] A. Brenier, G. Jia, and Ch. Tu, J. Phys.: Condens. Matter **16**, 9103 (2004).

[14] D. Errandonea, M. Somayazulu, and D. Häusermann, phys. stat. sol. (b) **235**, 162 (2003).

[15] D. Errandonea, M. Somayazulu, and D. Häusermann, phys. stat. sol. (b) **231**, R1 (2002).

[16] V. Panchal, N. Garg, A. K. Chauhan, Sangeeta, and S. M. Sharma, Solid State Commun. **130**, 203 (2004).

[17] A. Grzechnik, W. A. Crichton, M. Hanfland, and S. Van Smaalen, J. Phys.: Condens. Matter **15**, 7261 (2003).

[18] D. Errandonea, F. J. Manjón, M. Somayazulu, and D. Häusermann, J. Solid Stat. Chem. **177**, 1087 (2004).

[19] R. M. Hazen, L. W. Finger, and J. W. E. Mariathasan, J. Phys. Chem. Solids **46**, 253 (1985).

[20] D. Errandonea, J. Pellicer-Porres, F. J. Manjón, A. Segura, Ch. Ferrer-Roca, R. S. Kumar, O. Tschauner, P. Rodriguez-Hernandez, J. Lopez-Solano, A. Mujica, A. Muñoz, and G. Aquilanti, Phys. Rev. B **72**, 174106 (2005).

[21] P. Rodríguez-Hernández, J. López-Solano, S. Radescu, A. Mujica, A. Muñoz, D. Errandonea, J. Pellicer-Porres, A. Segura, Ch. Ferrer-Roca, F. J. Manjón, R. S. Kumar, O. Tschauner, and G. Aquilanti, to appear in J. Phys. Chem. Solids (2006).

[22] D. Errandonea, phys. stat. sol. (b) **242**, R125 (2005).

[23] A. Grzechnik, W. A. Crichton, and M. Hanfland, phys. stat. sol. (b) **242**, 2795 (2005).

[24] T. Fujita, S. Yamaoka, and O. Fukunaga, Mater. Res. Bull. **9**, 141 (1974).

[25] I. Kawada, K. Kato, and T. Fujita, Acta Cryst. B **30**, 2069 (1974).

[26] P. W. Richter, G. J. Kruger, and C. W. F. T. Pistorius, Acta Cryst. B **32**, 928 (1976).

[27] L. L. Y. Chang, J. Am. Ceram. Soc. **54**, 357 (1971).

[28] T. Fujita, I. Kawada, and K. Kato, Acta Cryst. B **33**, 162 (1977).

[29] A. Jayaraman, B. Batlogg, and L. G. Van Uitert, Phys. Rev. B **28**, 4774 (1983).

[30] A. Jayaraman, B. Batlogg, and L. G. Van Uitert, Phys. Rev. B **31**, 5423 (1985).

[31] W. Jeitschko and A.W. Sleight. Acta Cryst. B**29**, 869 (1973).

[32] S. Li, R. Ahuja, Y. Wang, and B. Johansson, High Press. Res. **23**, 343 (2003).

[33] H. K. Mao, J. Xu, and P. M. Bell, J. Geophys. Res. **91**, 4673 (1986).

[34] A. P. Hammersley, S. O. Svensson, M. Hanfland, A. N. Fitch, and D. Häusermann. High Pres. Res. **14**, 235 (1996).

[35] A. C. Larson and R. B. Von Dreele, LANL Report 86-748 (2004).

[36] W. Kraus and G. Nolze, J. Appl. Crystallogr. **29**, 301 (1996).

[37] M. Hagelstein, A. San Miguel, A. Fontaine, and J. Goulon, J. Phys. IV **7**, 303 (1997).

[38] S. Pascarelli, O. Mathon, and G. Aquilanti, J. Alloys Compds. **362**, 33 (2004).

[39] J. Pellicer-Porres, A. San Miguel, and A. Fontaine, J. Synchrotron Radiat. **5**, 1250 (1998).

[40] G. Kresse and J. Furthmüller, Comput. Mat. Sci. 6, 15-50 (1996), Phys. Rev. B **54**, 11169 (1996).

[41] J. P. Perdew, J. A. Chevary, S. H. Vosko, K. A. Jackson, M. R. Pederson, and D. J. Singh, and C. Fiolhais. Phys. Rev. B **46**, 6671 (1992).

[42] A. Mujica, A. Rubio, A. Muñoz, and R. J. Needs, Rev. Mod. Phys. **75**, 863 (2003).

[43] D. Vanderbilt, Phys. Rev. B **41**, 7892 (1990); G. Kresse and J. Hafner, J. Phys.: Condens. Matter **6**, 8245 (1994).

[44] P.E. Blöchl, Phys. Rev. B **50**, 17953 (1994); G. Kresse and D. Joubert, Phys. Rev. B **59**, 1758 (1999).

[45] J. M. Besson, J. P. Itie, A. Polian, G. Weill, J. L. Mansot, and J. Gonzalez, Phys. Rev. B **44**, 4214 (1991).

[46] A. Le Bail, H. Duray, and J. L. Fourquet, Mater. Res. Bull. **23**, 447 (1985).

[47] M. Daturi, M. M. Borel, A. Leclaire, L. Savary, G. Costentin, J. C. Lavalley, and B. Raveau, J. Chem. Phys. Phys.-Chem. Biol. **93**, 2043 (1996).

[48] Y. A. Titov, A. M. Sych, A. N. Sokolov, A. A. Kapshuk, V. Ya Markiv, and N. M. Belyavina, J. Alloys Compds. **311**, 252 (2000).

[49] E. T. Keve, S. C. Abrahams, and J. L. Bernstein, J. Chem. Phys. **51**, 4928 (1969).

[50] B. O. Loopstra and H. M. Rietveld, Acta Cryst. B **25**, 787 (1969).

[51] R. W. G. Wyckoff, Zeitschrift fur Kristallographie **66**, 73 (1927).

[52] J. Beintema, Zeitschrift fur Kristallographie **97**, 300 (1937).

[53] D. Errandonea, Y. Meng, M. Somayazulu, and D. Häusermann, Physica B **355**, 116 (2005), and references therein.

[54] O. Tschauner, D. Errandonea, and G. Serghiou, Physica B **371**, 88 (2006).

[55] A.Grzechnik, W. A. Crichton, P. Bouvier, V. Dmitriev, H. P. Weber, and J. Y. Gesland, J. Phys.: Condens. Matter **16**, 7779 (2004).

[56] J. Ankudinov, B. Ravel, and S. Conradson, Phys. Rev. B **58**, 7565 (1998).

[57] *High-pressure lattice dynamics in bulk single-crystal $PbWO_4$,* F. J. Manjon, D. Errandonea, N. Garro, J. Pellicer-Porres, J. Lopez-Solano, P. Rodriguez-Hernandez, and A. Muñoz, cond-mat/060547 (2006).

[58] *Effects of high pressure on the optical absorption spectrum of scintillating $PbWO_4$ crystals*, D. Errandonea, D. Martínez-García, R. Lacomba-Perales, J. Ruiz-Fuertes, and A. Segura, cond-mat/060549 (2006).

[59] R.J. Cava and R.S. Roth, J. Solid State Chem. **36**, 139 (1981).

[60] J. Lopez-Solano, P. Rodríguez-Hernández, A. Muñoz, and F.J. Manjón, Phys. Rev. B **73**, 094117 (2006).

[61] E. Knittle and Q. Williams, Am. Mineral. **78**, 245 (1993).

[62] V. P. Prakapenka, G. Shen, L. S. Dubrovinsky, M. L. Rivers, and S. R. Sutton, J. Phys. Chem. Solids **65**, 1537 (2004).

[63] T. Sasaki, J. Phys.: Condens. Matter **14**, 10557 (2002).

[64] J. F. Kenney, V. A. Kutcherov, N. A. Bendeliani, and V. A. Alekseev, *Proc. Natl. Acad. Sci. USA* **99,** 10976 (2002).

**Table I:** Bulk modules ($B_0$), first pressure derivate of the bulk modulus ($B_0'$), and equilibrium volume ($V_0$) all at normal conditions for different $AWO_4$ compounds. [a]Ref. [20], [b]Ref. [22], and [c]present work.

| Compound | $V_0$ [Å$^3$] | $B_0$ [GPa] | $B_0'$ |
|---|---|---|---|
| $CaWO_4$[a] | 312(1) | 74(7) | 5.6(9) |
| $SrWO_4$[a] | 347.4(9) | 63(7) | 5.2(9) |
| $BaWO_4$[c] | 402.8(9) | 52(5) | 5(1) |
| $PbWO_4$[c] | 357.8(6) | 66(5) | 5.6(9) |
| $EuWO_4$[b] | 348.9(8) | 65(6) | 4.6(9) |

**Table II:** Structural parameters of the different structures of $BaWO_4$ and $PbWO_4$ as obtained from the present Rietveld refinements.

Structural parameters of scheelite $BaWO_4$ at 1 GPa:
$I4_1/a$, Z = 4, $a$ = 5.603(4) Å, $c$ = 12.693(7) Å

|    | Site | x         | y         | z         |
|----|------|-----------|-----------|-----------|
| Ba | 4b   | 0         | 0.25      | 0.625     |
| W  | 4a   | 0         | 0.25      | 0.125     |
| O  | 16f  | 0.2336(4) | 0.0976(6) | 0.0499(6) |

Structural parameters of scheelite $BaWO_4$ at 6.9 GPa:
$I4_1/a$, Z = 4, $a$ = 5.460(4) Å, $c$ = 12.138(7) Å

|    | Site | x         | y         | z         |
|----|------|-----------|-----------|-----------|
| Ba | 4b   | 0         | 0.25      | 0.625     |
| W  | 4a   | 0         | 0.25      | 0.125     |
| O  | 16f  | 0.2222(4) | 0.1087(6) | 0.0499(6) |

Structural parameters of fergusonite $BaWO_4$ at 7.3 GPa:
$I2/a$, Z = 4, $a$ = 5.465(7) Å, $b$ = 12.109(3) Å, $c$ = 5.439(7) Å, $\beta$ = 90.087(9)°

|       | Site | x          | y          | z          |
|-------|------|------------|------------|------------|
| Ba    | 4e   | 0.25       | 0.6131(9)  | 0          |
| W     | 4e   | 0.25       | 0.1282(7)  | 0          |
| $O_1$ | 8f   | 0.9493(49) | 0.9765(23) | 0.2538(34) |
| $O_2$ | 8f   | 0.4666(36) | 0.2225(37) | 0.8826(37) |

Structural parameters of scheelite $PbWO_4$ at 0.7 GPa:
$I4_1/a$, Z = 4, $a$ = 5.436(7) Å, $c$ = 11.957(9) Å

|    | Site | x         | y         | z         |
|----|------|-----------|-----------|-----------|
| Pb | 4b   | 0         | 0.25      | 0.625     |
| W  | 4a   | 0         | 0.25      | 0.125     |
| O  | 16f  | 0.2482(6) | 0.1073(6) | 0.0510(9) |

**Table III:** Calculated structural parameters of different structures of BaWO$_4$ and PbWO$_4$.

BaWO$_4$ in the BaMnF$_4$-type structure at 9.1 GPa:

A2$_1$/*am*, Z = 4, *a* = 5.8252 Å, *b* = 13.8973 Å, *c* = 3.9528 Å

|       | Site | x      | y      | z |
|-------|------|--------|--------|---|
| Ba    | 4a   | 0.9959 | 0.1315 | 0 |
| W     | 4a   | 0.4962 | 0.6015 | 0 |
| O$_1$ | 4a   | 0.7240 | 0.6896 | 0 |
| O$_2$ | 4a   | 0.2681 | 0.6894 | 0 |
| O$_3$ | 4a   | 0.7470 | 0.5002 | 0 |
| O$_4$ | 4a   | 0.4565 | 0.0862 | 0 |

BaWO$_4$ in the LaTaO$_4$-type structure at 8.9 GPa:

P2$_1$/*c*, Z = 4, *a* = 7.2255 Å, *b* = 5.7858 Å, *c* = 8.0620 Å, β = 106º

|       | Site | x      | y      | z      |
|-------|------|--------|--------|--------|
| Ba    | 4e   | 0.2707 | 0.7500 | 0.0760 |
| W     | 4e   | 0.2003 | 0.2500 | 0.3241 |
| O$_1$ | 4e   | 0.1721 | 0.2500 | 0.0466 |
| O$_2$ | 4e   | 0.0004 | 0.5007 | 0.2502 |
| O$_3$ | 4e   | 0.3769 | 0.4781 | 0.3472 |
| O$_4$ | 4e   | 0.3767 | 0.0216 | 0.3472 |

BaWO$_4$-II phase at 9.3 GPa:
P2$_1$/*n*, Z = 8, a = 12.7173 Å, b = 6.9816 Å, c = 7.4357 Å, β = 91.22º

|         | Site | x      | y      | z      |
|---------|------|--------|--------|--------|
| Ba$_1$  | 4e   | 0.1617 | 0.6555 | 0.1633 |
| Ba$_2$  | 4e   | 0.1349 | 0.9574 | 0.6316 |
| W$_1$   | 4e   | 0.0825 | 0.1633 | 0.0836 |
| W$_2$   | 4e   | 0.0912 | 0.4609 | 0.6497 |
| O$_1$   | 4e   | 0.1078 | 0.0279 | 0.2876 |
| O$_2$   | 4e   | 0.1845 | 0.6029 | 0.7777 |
| O$_3$   | 4e   | 0.0490 | 0.6510 | 0.4746 |
| O$_4$   | 4e   | 0.2128 | 0.2676 | 0.0618 |
| O$_5$   | 4e   | 0.0579 | 0.2693 | 0.8168 |
| O$_6$   | 4e   | 0.1783 | 0.3319 | 0.5103 |
| O$_7$   | 4e   | 0.0198 | 0.3756 | 0.1829 |
| O$_8$   | 4e   | 0.0789 | 0.9201 | 0.9539 |

Fergusonite PbWO$_4$ at 7.9 GPa:
I2/*a*, Z = 4, *a* = 5.415 Å, *b* = 11.661 Å, *c* = 5.385 Å, β = 90.11°

|       | Site | x      | y      | z      |
|-------|------|--------|--------|--------|
| Pb    | 4e   | 0.25   | 0.6246 | 0      |
| W     | 4e   | 0.25   | 0.1263 | 0      |
| O$_1$ | 8f   | 0.9021 | 0.9598 | 0.2308 |
| O$_2$ | 8f   | 0.4806 | 0.2113 | 0.8455 |

Ferguson ite PbWO$_4$ at 9.5 GPa:
I2/a, Z = 4, a = 5.900 Å, b = 11.090 Å, c = 4.923 Å, β = 96.601°

|       | Site | x      | y      | z      |
|-------|------|--------|--------|--------|
| Pb    | 4e   | 0.25   | 0.6175 | 0      |
| W     | 4e   | 0.25   | 0.1521 | 0      |
| O$_1$ | 8f   | 0.9049 | 0.9554 | 0.2246 |
| O$_2$ | 8f   | 0.4528 | 0.2126 | 0.7674 |

PbWO$_4$ in the LaTaO$_4$-type structure at 10 GPa:
P 2$_1$/c, Z =4, a = 7.4233 Å, b = 5.6241 Å, c = 7.9757 Å, β = 106.5°

|       | Site | x      | y      | z      |
|-------|------|--------|--------|--------|
| Pb    | 4e   | 0.3355 | 0.7593 | 0.0718 |
| W     | 4e   | 0.1809 | 0.2521 | 0.2634 |
| O$_1$ | 4e   | 0.1688 | 0.2226 | 0.0346 |
| O$_2$ | 4e   | 0.0186 | 0.5338 | 0.2390 |
| O$_3$ | 4e   | 0.3765 | 0.4768 | 0.3334 |
| O$_4$ | 4e   | 0.3580 | 0.0169 | 0.3456 |

PbWO$_4$-III phase at 12 GPa:
P2$_1$/n, Z = 8, a = 12.2171 Å, b = 6.8146 Å, c = 7.2069 Å, β = 89.63°

|        | Site | x      | y      | z      |
|--------|------|--------|--------|--------|
| Pb$_1$ | 4e   | 0.1591 | 0.6760 | 0.1645 |
| Pb$_2$ | 4e   | 0.1356 | 0.9576 | 0.6238 |
| W$_1$  | 4e   | 0.0850 | 0.1680 | 0.0824 |
| W$_2$  | 4e   | 0.0968 | 0.4637 | 0.6444 |
| O$_1$  | 4e   | 0.1036 | 0.0197 | 0.2883 |
| O$_2$  | 4e   | 0.1941 | 0.6057 | 0.7748 |
| O$_3$  | 4e   | 0.0474 | 0.6587 | 0.4739 |
| O$_4$  | 4e   | 0.2228 | 0.2676 | 0.0661 |
| O$_5$  | 4e   | 0.0663 | 0.2606 | 0.8137 |
| O$_6$  | 4e   | 0.1836 | 0.3375 | 0.4863 |
| O$_7$  | 4e   | 0.0209 | 0.3894 | 0.1801 |
| O$_8$  | 4e   | 0.0824 | 0.9126 | 0.9489 |

PbWO$_4$ in the Cmca-type structure at 11.7 GPa:
Cmca, Z = 8, a = 7.8152 Å, b = 13.3597 Å, c = 5.3635 Å

|       | Site | x      | y      | z      |
|-------|------|--------|--------|--------|
| Pb    | 8e   | 0.25   | 0.8313 | 0.25   |
| W     | 8f   | 0.5    | 0.5859 | 0.2841 |
| O$_1$ | 8e   | 0.25   | 0.1352 | 0.25   |
| O$_2$ | 8f   | 0.5    | 0.7977 | 0.9619 |
| O$_3$ | 8d   | 0.1450 | 0      | 0.5    |
| O$_4$ | 8f   | 0.5    | 0.9134 | 0.4396 |

**Figure captions**

**Fig. 1.** Room-temperature ADXRD data of (a) BaWO$_4$ and (b) PbWO$_4$ at different pressures. In all diagrams the background was subtracted. To better illustrate the appearance of the (020) Bragg reflection of the fergusonite structure around 2θ ≈ 4º a section of one of the fergusonite patterns is enlarged and shown in the inset. In (b) to illustrate the quality of the Le Bail refinement obtained for the fergusonite structure at 10.1 GPa the difference between the measured data and the refined profile is shown (dotted line). The bars indicate the calculated positions of the reflections of the fergusonite structure.

**Fig. 2.** Evolution of the lattice parameters (a) and (b), volume (c), and axial ratios (d) of BaWO$_4$ and PbWO$_4$ under pressure. Empty squares correspond to data of the scheelite phase and empty (gray) circles to data of the fergusonite phase of BaWO$_4$ (PbWO$_4$). Solid circles **[19]**, solid triangles **[16]**, and crosses **[7]** illustrate the data of the scheelite phase obtained from the literature. Empty triangles are the fergusonite data reported in Ref. **[16]**. In (c) the solid lines represent the EOS described in the text. In (a), (b), and (d) they are just a guide to the eye. The plotted EOS corresponds to the true compressibility of the scheelite phase. However, it represents very well also the pressure dependence of the volume of the fergusonite phase.

**Fig. 3.** Pressure dependence of the interatomic bond distances in the scheelite and fergusonite phases of BaWO$_4$ and PbWO$_4$. Squares represent the calculated distances and circles the distances extracted from the present experiments.

**Fig. 4.** ADXRD pattern of BaWO$_4$ at 7.3 GPa and 10.9 GPa. The background was subtracted. The dotted lines represent the difference between the measured data and the refined profiles. The bars indicate the calculated positions of the reflections.

**Fig. 5.** *Ab initio* simulation of XANES spectra at the W L$_3$-edge of BaWO$_4$ (a) and PbWO$_4$ (b) in several different phases. The structural data used in simulations are taken from Tables II and III, and from references **[26, 31, and 47]** (see text for details). Resonance B is present when the W coordination is fourfold, but it is not present when the W coordination is sixfold.

**Fig. 6.** Experimental XANES spectra (W L$_3$-edge) of BaWO$_4$ (a) and PbWO$_4$ (b) at different pressures. The spectra collected on pressure release are marked with d. The analysis of the spectra (see text) reveals a coordination change at 9.8 GPa in the case of BaWO$_4$. In its turn, PbWO$_4$ transits to a sixfold coordinated fergusonite between 9 and 10.5 GPa. Subtle changes in the spectra suggest a second phase transition in PbWO$_4$ at 16.7 GPa.

**Fig. 7.** Total-energy versus volume *ab initio* calculations for BaWO$_4$ (a) and PbWO$_4$ (b) in different structures. The insets extend the pressure range to the region where the BaWO$_4$-II and PbWO$_4$-III structures becomes unstable.

**Figure 1a**

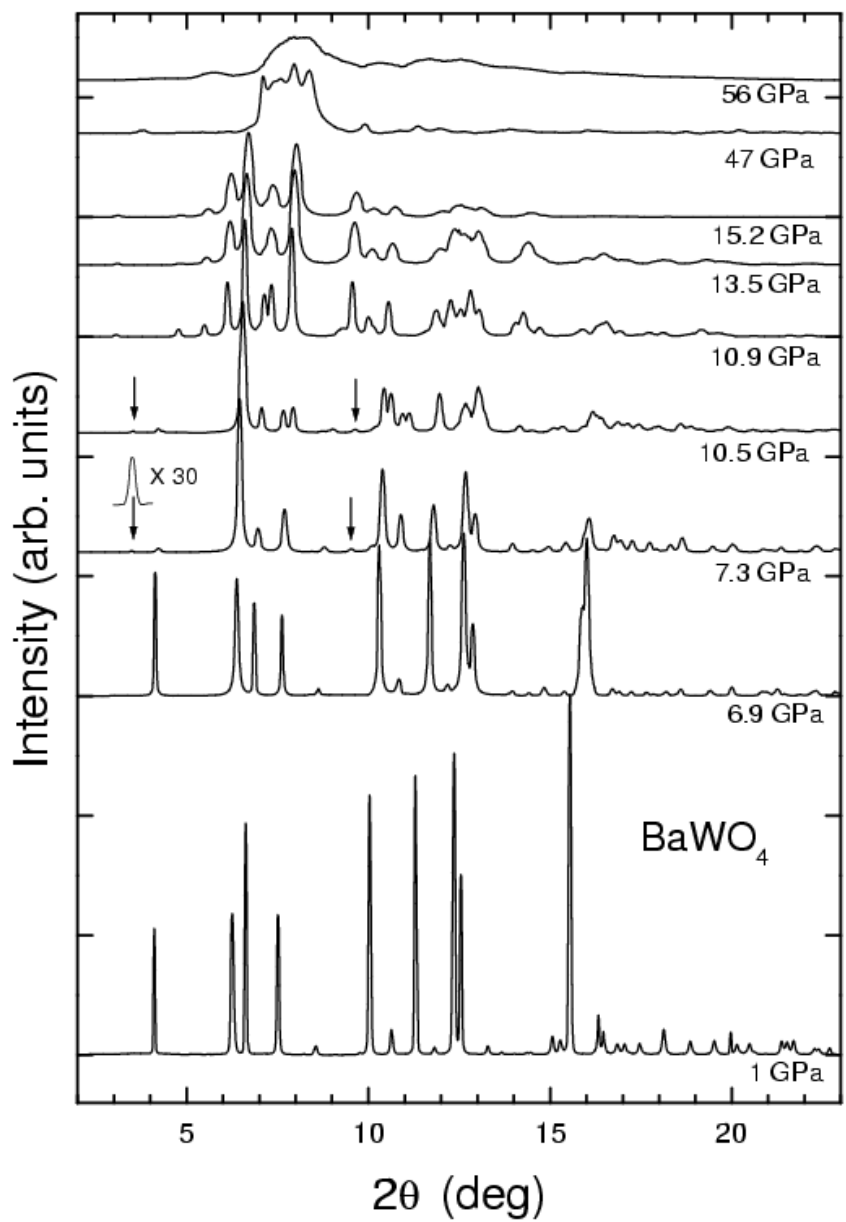

**Figura 1b**

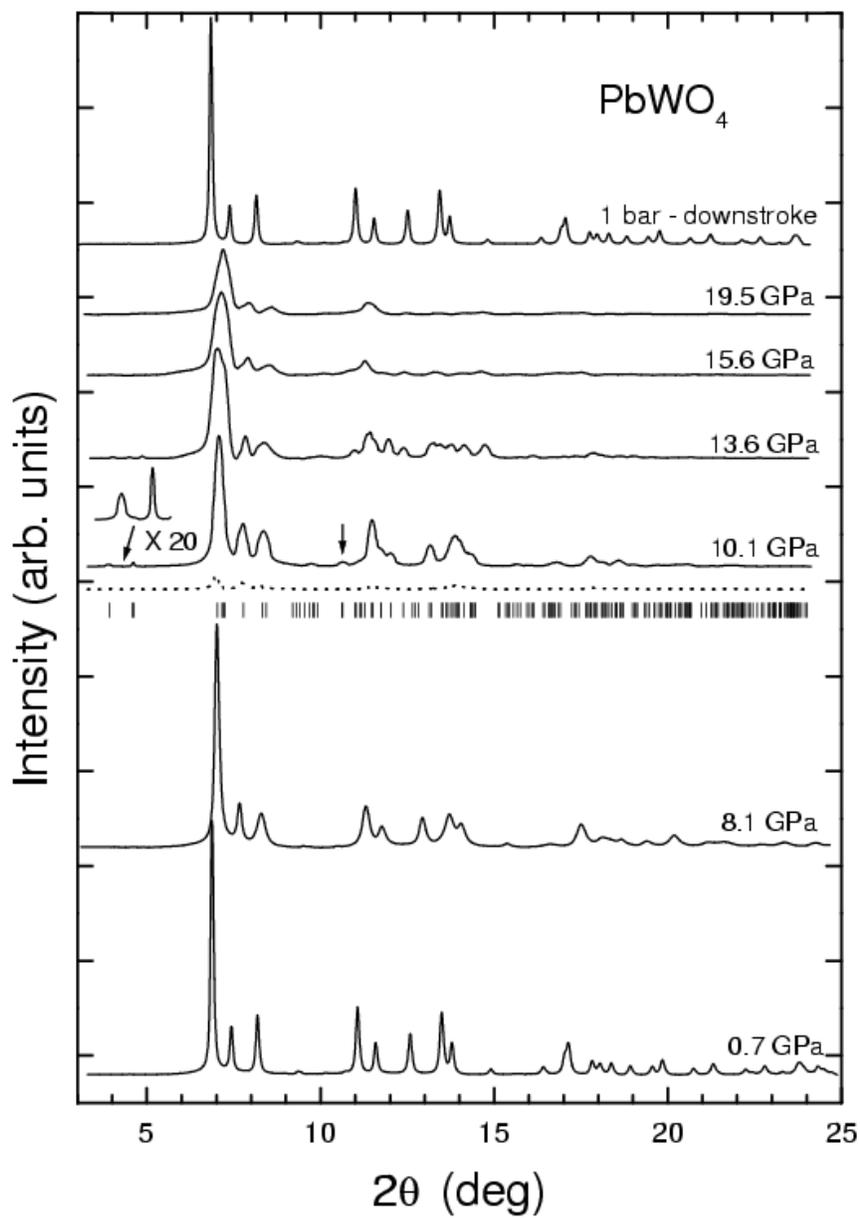

**Figure 2**

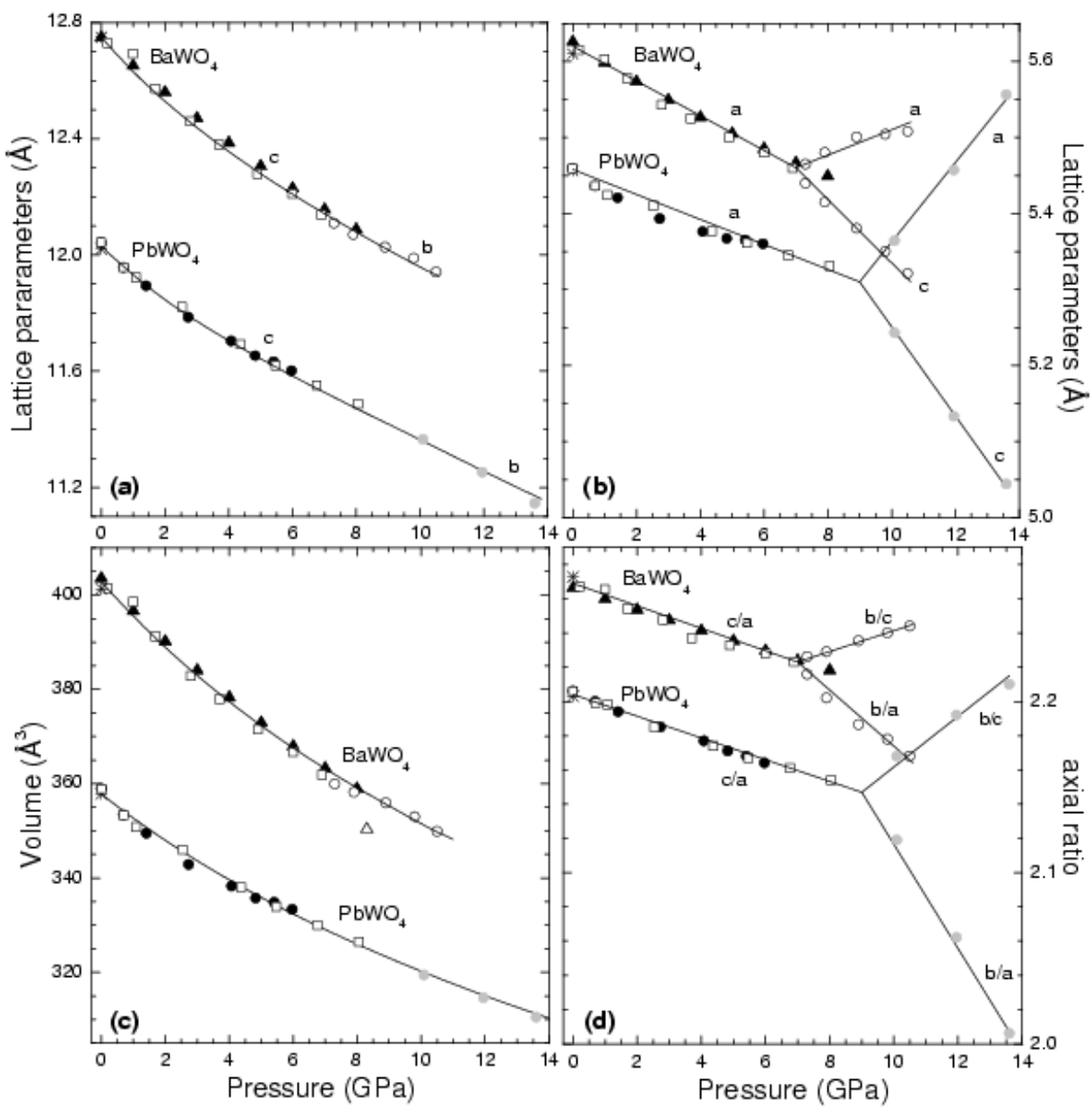

**Figure 3**

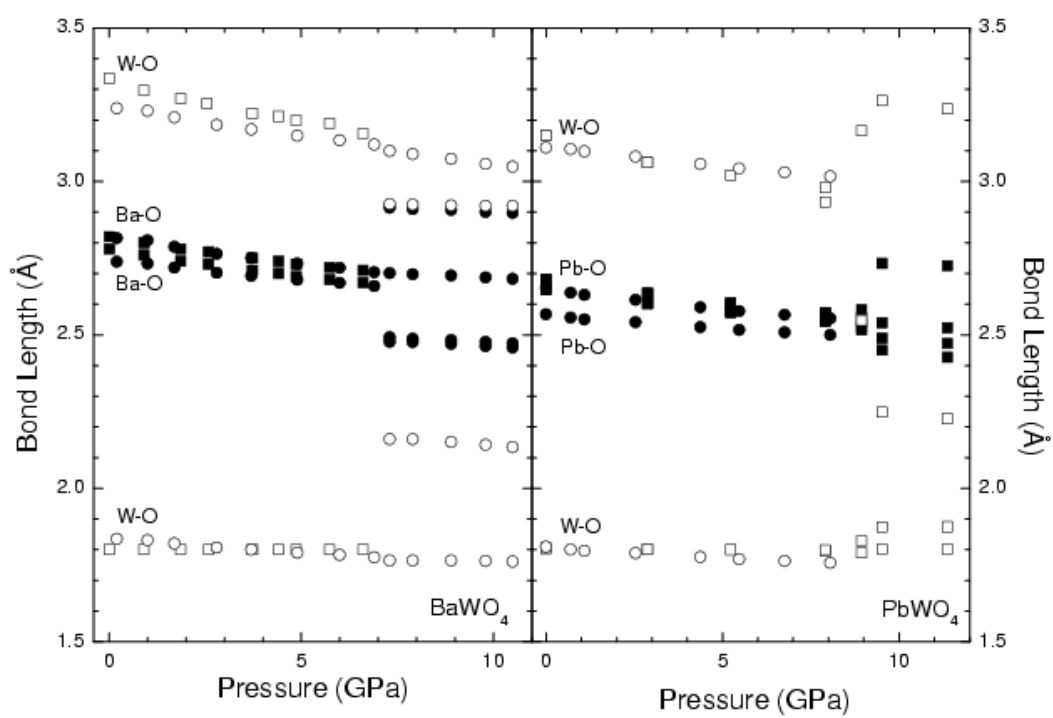

**Figure 4**

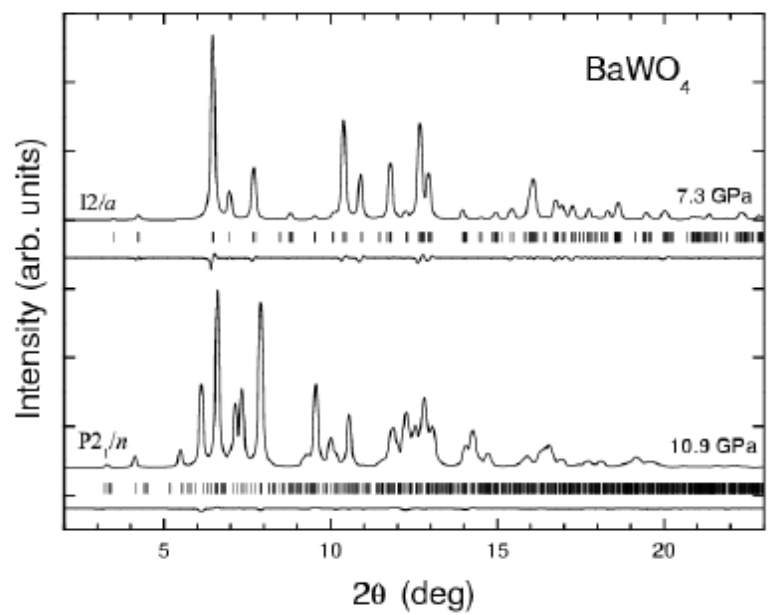

**Figure 5a**

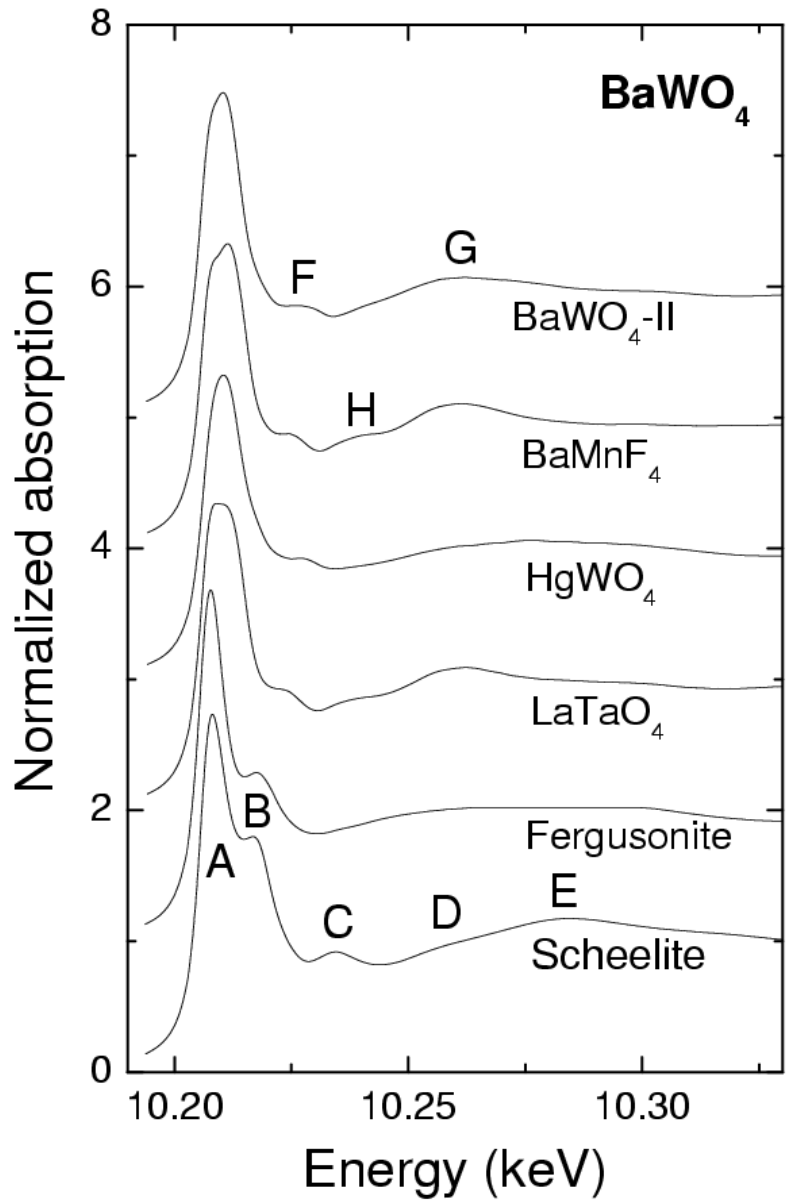

**Figure 5b**

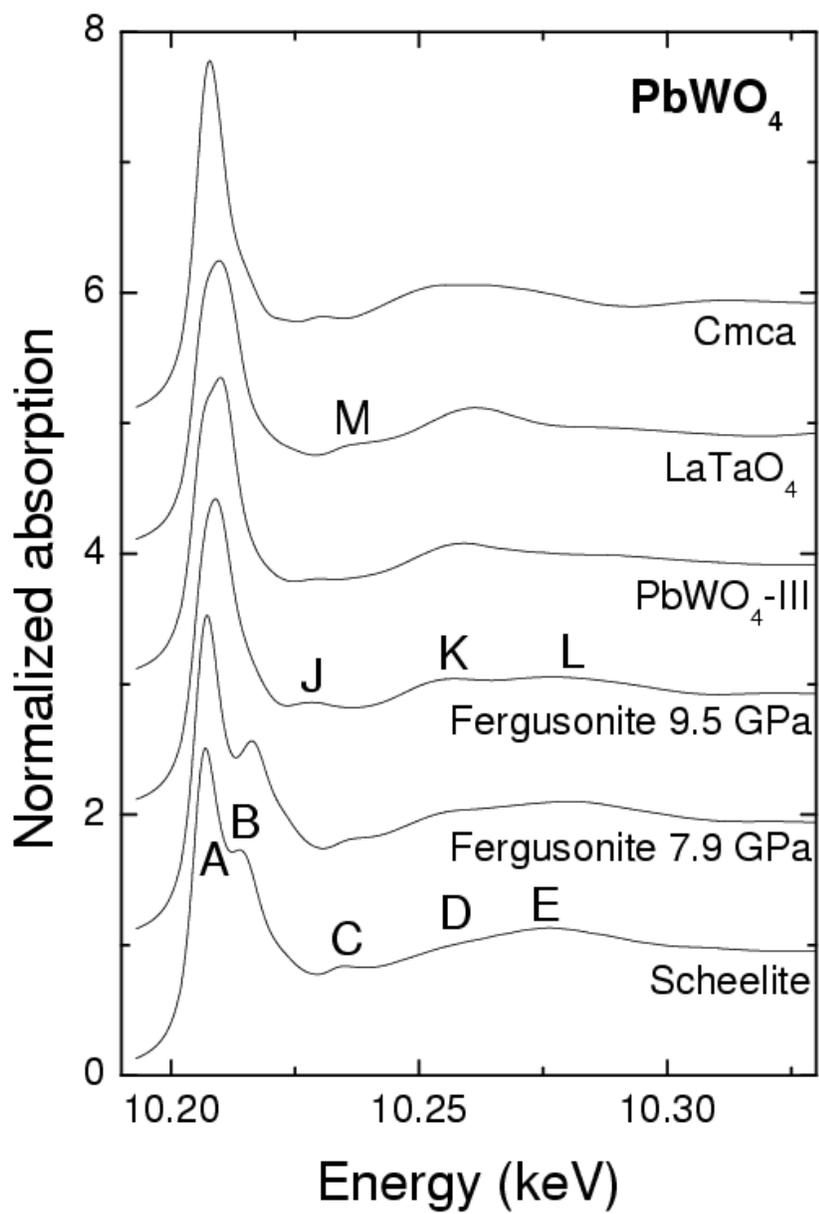

**Figure 6a**

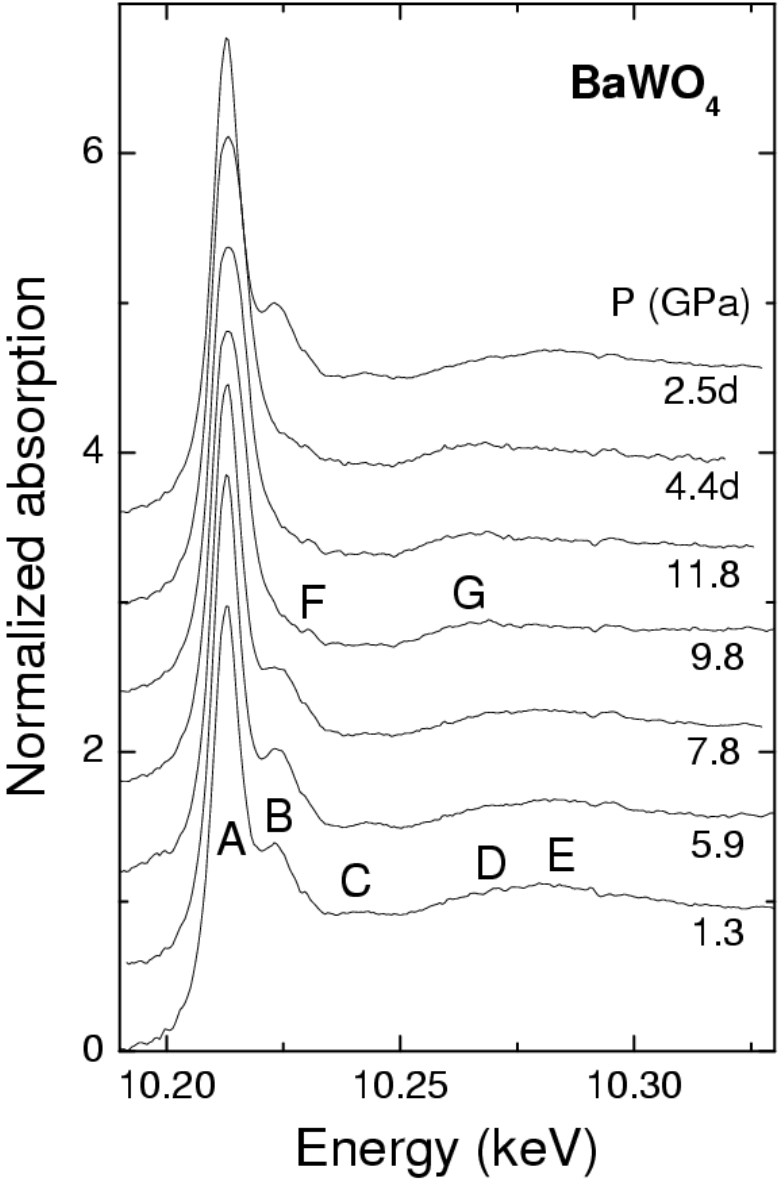

**Figure 6b**

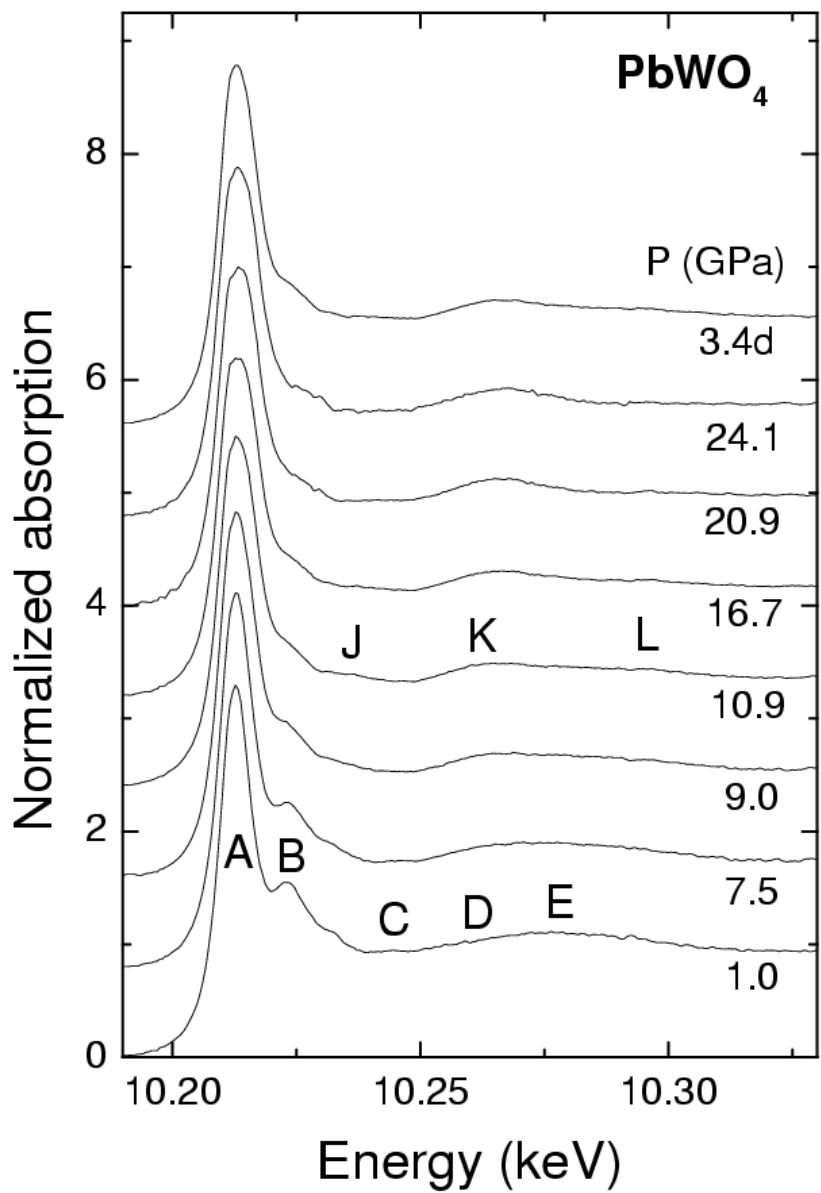

**Figure 7a**

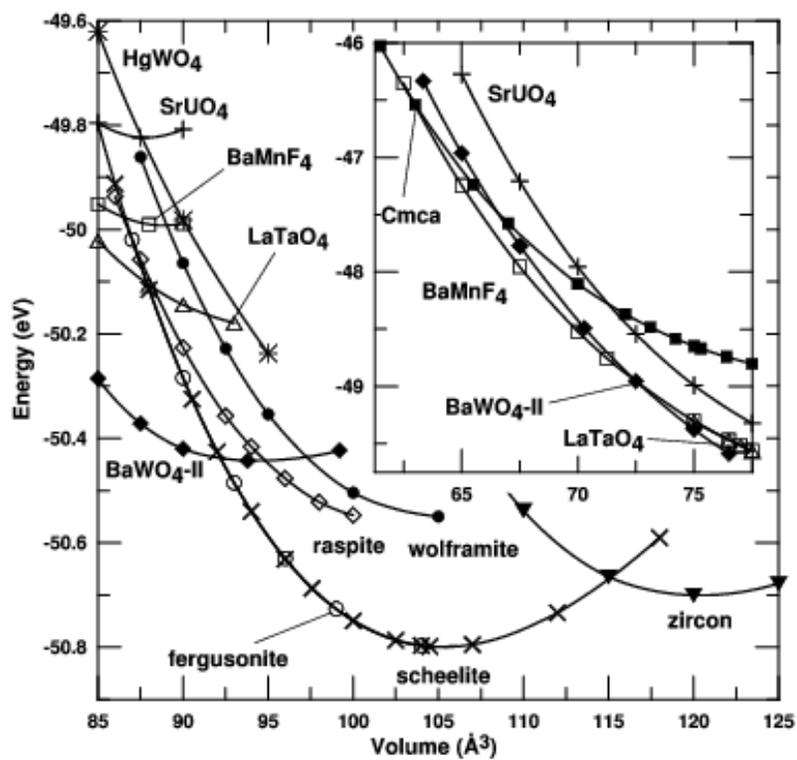

**Figure 7b**

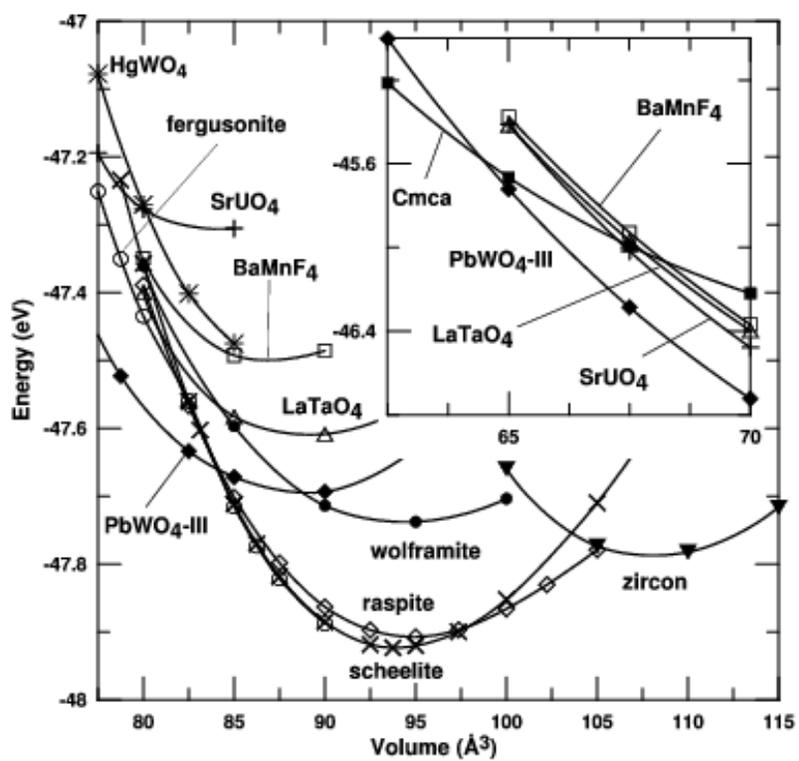